\documentclass[a4paper,11pt]{article}
\usepackage{pos}

\title{Search for the muon electric dipole moment using frozen-spin technique at PSI}

\author*[a,b]{Kim Siang Khaw}
\author[c,d]{Andreas Adelmann}
\author[c]{Malte Backhaus}
\author[e]{Niklaus Berger}
\author[d]{Manfred Daum}
\author[f]{Massimo Giovannozzi}
\author[c,d]{Klaus Kirch}
\author[d]{Andreas Knecht}
\author[d,g]{Angela Papa}
\author[d]{Claude Petitjean}
\author[h]{Francesco Renga}
\author[c]{Mikio Sakurai}
\author[d]{Philipp Schmidt-Wellenburg}

\affiliation[]{on behalf of the muon EDM initiative/collaboration}
\affiliation[a]{Tsung-Dao Lee Institute, Shanghai Jiao Tong University, Shanghai, China}
\affiliation[b]{School of Physics and Astronomy, Shanghai Jiao Tong University, Shanghai, China}
\affiliation[c]{Institute for Particle Physics and Astrophysics, ETH Zürich, 8093 Zürich, Switzerland}
\affiliation[d]{Paul Scherrer Institute, 5232 Viligen PSI, Switzerland}
\affiliation[e]{PRISMA$^{+}$ Cluster of Excellence and Institute of Nuclear Physics, Johannes Gutenberg University Mainz, Mainz, Germany}
\affiliation[f]{CERN, 1211 Geneva, Switzerland}
\affiliation[g]{Dipartimento di Fisica, University of Pisa and INFN Sezione di Pisa, Largo B. Pontecorvo 3, 56127 Pisa, Italy}
\affiliation[h]{Istituto Nazionale di Fisica Nucleare, Sez. di Roma, P.le A. Moro 2, 00185 Roma, Italy}

\emailAdd{kimsiang84@sjtu.edu.cn}

\abstract{
The presence of a permanent electric dipole moment in an elementary particle implies Charge-Parity symmetry violation and thus could help explain the matter-antimatter asymmetry observed in our universe. Within the context of the Standard Model, the electric dipole moment of elementary particles is extremely small. However, many Standard Model extensions such as supersymmetry predict large electric dipole moments. Recently, the muon electric dipole moment has become a topic of particular interest due to the tensions in the magnetic anomaly of the muon and the electron, and hints of lepton-flavor universality violation in B-meson decays. In this article, we discuss a dedicated effort at the Paul Scherrer Institute in Switzerland to search for the muon electric dipole moment using a 3-T compact solenoid storage ring and the frozen-spin technique. This technique could reach a sensitivity of $6\times10^{-23}$ $e\cdot$cm after a year of data taking with the $p=125$ MeV/$c$ muon beam at the Paul Scherrer Institute. This allows us to probe various Standard Model extensions not reachable by traditional searches using muon $g-2$ storage rings.}

\FullConference{%
  *** The 22nd International Workshop on Neutrinos from Accelerators (NuFact2021) ***\\
  *** 6–11 Sep 2021 ***\\
  *** Cagliari, Italy ***}

\begin{document}
\maketitle

\section{Introduction}
The muon electric dipole moment (EDM) is one of the least tested areas of the Standard Model (SM) of particle physics. The current experimental limit of $d_{\mu} < 1.8 \times 10^{-19} \,e\cdot$cm at 95\% confidence level~\cite{Muong-2:2008ebm} from the BNL Muon ($g-2$) Collaboration is several orders of magnitude away from the SM prediction of $10^{-42}\,e\cdot$cm~\cite{Pospelov:2013sca}, assuming linear mass scaling $d_{\mu}/d_\mathrm{e} \sim m_{\mu}/m_\mathrm{e}$. A permanent muon EDM larger than the SM prediction, as anticipated by several models like supersymmetry, will be a new source of CP-violation beyond the SM, if found experimentally. Recently, the muon EDM has become a topic of particular interest~\cite{Crivellin:2018qmi} due to the tensions in the magnetic anomaly of the muon $a_{\mu}$~\cite{Aoyama:2020ynm, Muong-2:2021ojo} and the electron $a_\mathrm{e}$~\cite{Parker:2018vye, Morel:2020dww}, and hints of lepton-flavor universality violation in B meson decay~\cite{Crivellin:2021sff}.

The muon EDM can be defined as $\vec{d}_{\mu}=\eta \frac{e}{2m_{\mu}c}\vec{s}$, in the same way as the muon magnetic-dipole moment $\vec{\mu}_{\mu}=g_{\mu}\frac{e}{2m_{\mu}}\vec{s}$. Here, $\eta$ is a dimensionless parameter and $|\vec{d}_{\mu}| \approx \eta  \times 4.7 \times 10^{-14}$. When a spin-polarized muon is placed in a magnetic field $\vec{B}$ and an electric field $\vec{E}$, the angular precession frequency of the spin $\vec{\omega}_\mathrm{s}$ relative to the cyclotron frequency $\vec{\omega}_\mathrm{c}$ is given by
\begin{equation}
    \vec{\omega}_\mathrm{s} - \vec{\omega}_\mathrm{c} = -\dfrac{e}{m_{\mu}} \Big\{  a_{\mu}\vec{B} + \big( \dfrac{1}{\gamma^2-1} -a_{\mu}\big)   \dfrac{\vec{\beta} \times \vec{E}}{c} + \dfrac{\eta}{2} \big( \dfrac{\vec{E}}{c} + \vec{\beta} \times \vec{B}\big) \Big\}~.
\end{equation}

\begin{figure}[htbp]
\centering
\includegraphics[width=0.5\textwidth]{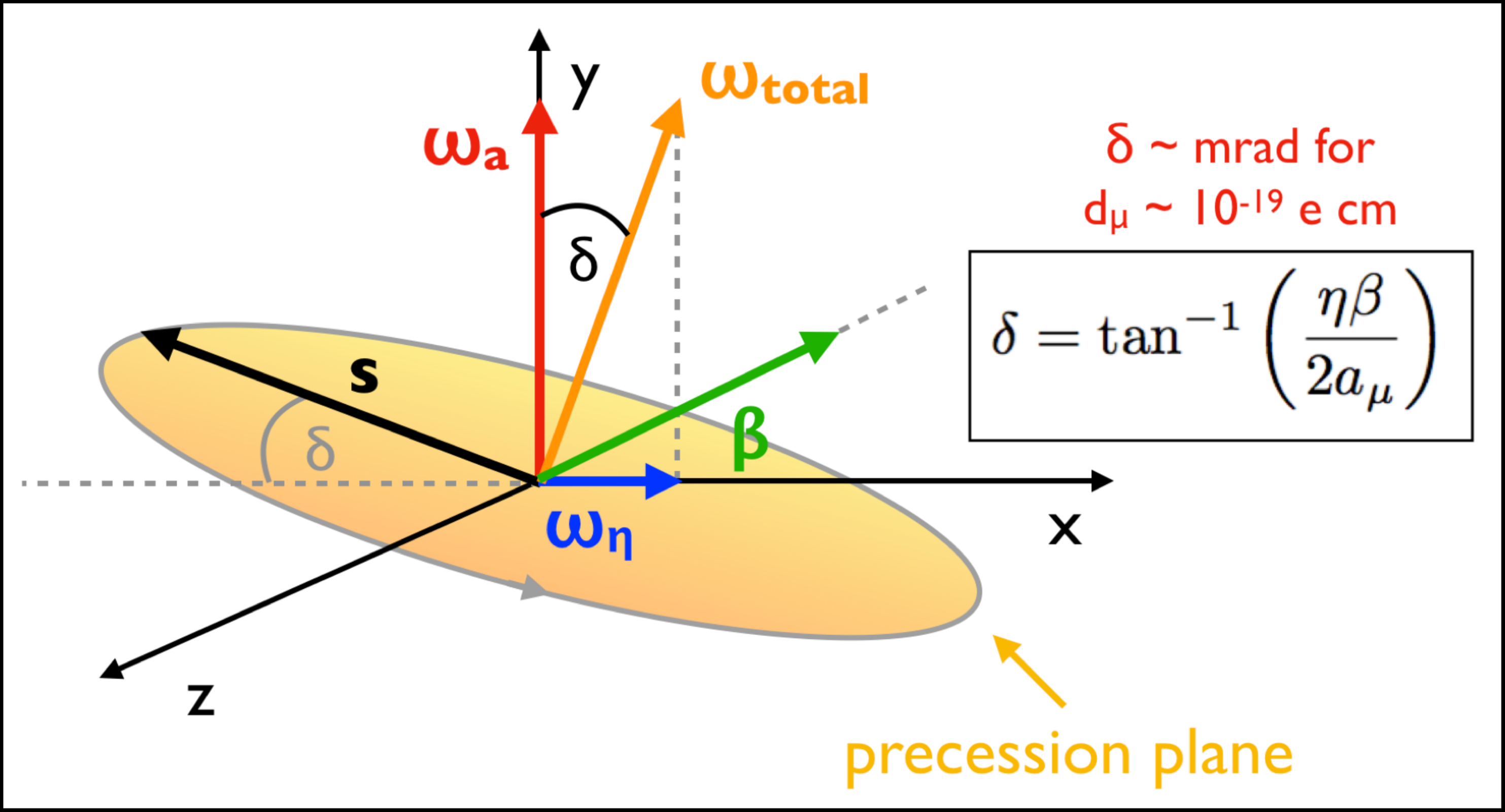}
\caption{Tilt of precession plane due to a non-zero muon EDM.}
\label{fig:tiltprecessionplane}
\end{figure}

A previous effort at BNL~\cite{Muong-2:2008ebm}, and current efforts at Fermilab~\cite{Chislett:2016jau} and at J-PARC~\cite{Abe:2019thb} searching for a muon EDM exploited the fact that in a muon $g-2$ storage-ring experiment, the existence of an EDM, i.e.\ when $\eta \ne 0$, will tilt the originally horizontal precession plane of the muon spin, as shown in Fig.~\ref{fig:tiltprecessionplane}. The current BNL limit gives a plane tilt $\delta$ of about mrad and this translates into an average vertical angle oscillation of several tens of $\mu$rad for the emitted positrons.  Improving further the sensitivity of this approach will be very challenging, due to multiple Coulomb scattering in tracking detectors and systematic effects such as beam-detector alignment and radial magnetic field. The projected sensitivity for the FNAL and J-PARC experiments is of the order of $10^{-21}\,e\cdot$cm~\cite{Chislett:2016jau, Abe:2019thb}. Recently, a \emph{frozen-spin} technique has been proposed~\cite{Farley:2003wt, Adelmann:2010zz}, which has the potential to reach a muon EDM sensitivity as low as $10^{-23}\,e\cdot$cm. Here we report on the status of an on-going effort at the Paul Scherrer Institute (PSI) utilizing the frozen-spin technique and a compact storage ring~\cite{Adelmann:2021udj}.

\section{A frozen-spin-based muon EDM search at PSI}

\begin{figure}[h]
\centering
\includegraphics[width=0.6\textwidth]{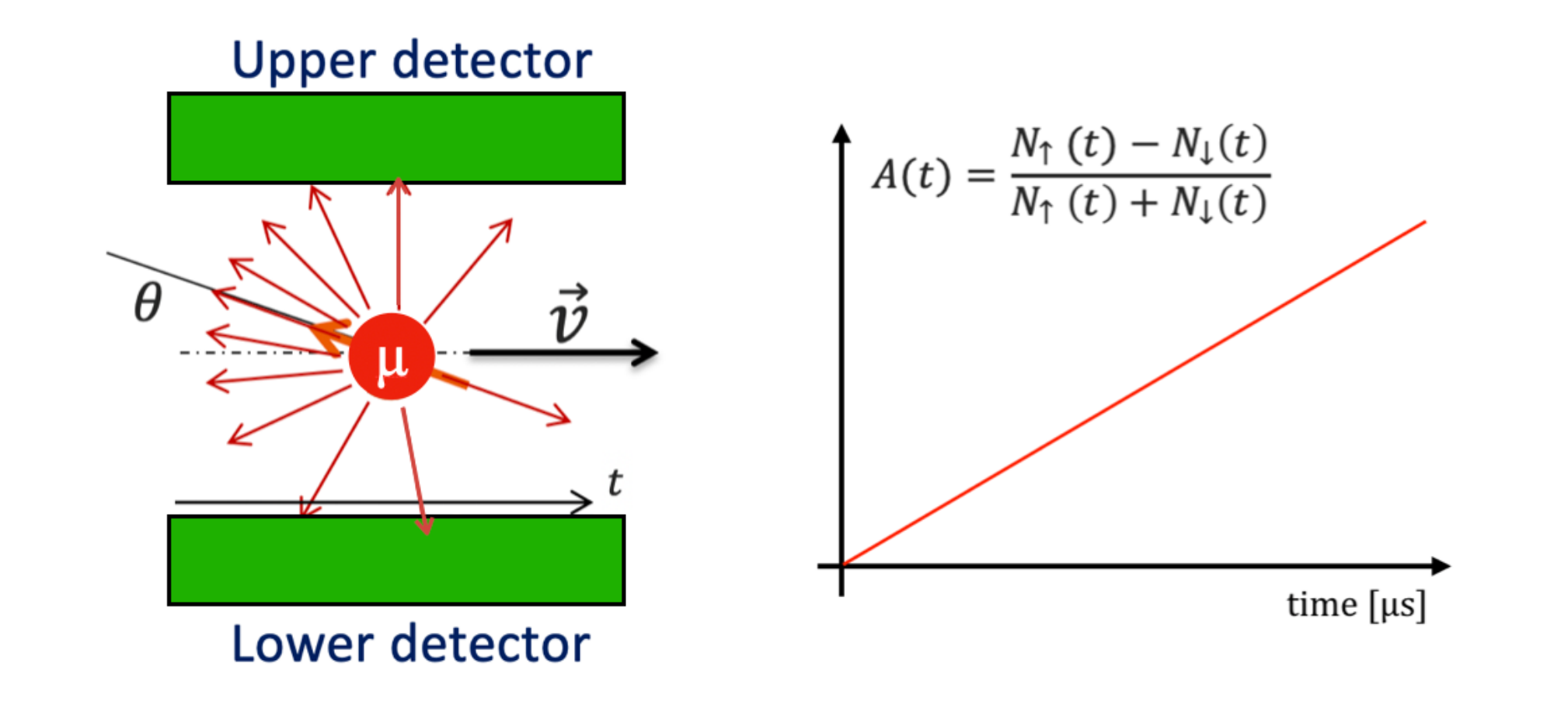}
\caption{Left: A sketch of the detection concept for the frozen-spin technique muon EDM search. Right: A slowly-increasing asymmetry versus time of the upper and lower detector due to a non-zero muon EDM.}
\label{fig:frozen-spin-asymmetry}
\end{figure}
For the frozen-spin technique, a radial electric field $E_{r} = aBc\beta\gamma^2 $ is applied to cancel the anomalous precession of the muon spin. As a result, the torque, due to the muon EDM, acting on the muon spin can cause an increase in the vertical polarization over time. By placing detectors at the top and bottom of the muon storage region, as the positron is emitted preferentially along the direction of the muon spin, one will observe an up-down asymmetry of the positron count due to the increasing polarization in the vertical direction as shown in Fig.~\ref{fig:frozen-spin-asymmetry}. The slope of the asymmetry versus time $A(t)$ then gives the sensitivity of the EDM measurement:
\begin{equation}
    \sigma(d_{\mu}) = \dfrac{\hbar\gamma^2 a_{\mu}}{2P E_\mathrm{f}\sqrt{N}\gamma\tau_{\mu}\alpha}
    \label{eq:edm-formula}
\end{equation}
where $P$ is the polarization, $E_\mathrm{f}$ is the electric field in the lab frame, $N$ is the number of positron, $\tau_{\mu}$ the muon lifetime, and $\alpha$ is the decay asymmetry.

\begin{figure}[htbp]
\centering
\includegraphics[width=0.45\textwidth]{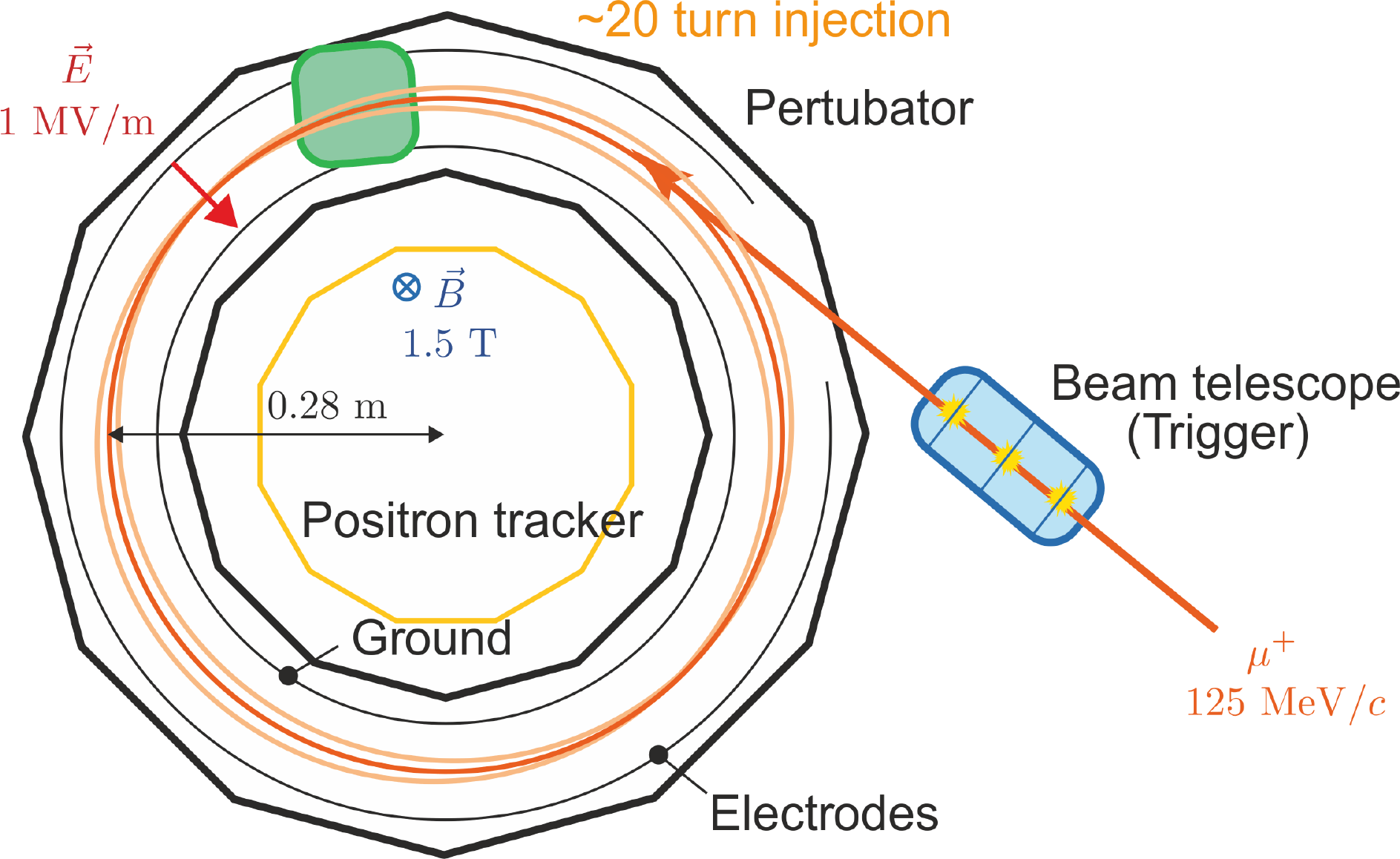}
\includegraphics[width=0.25\textwidth,trim=0 30 0 0, clip]{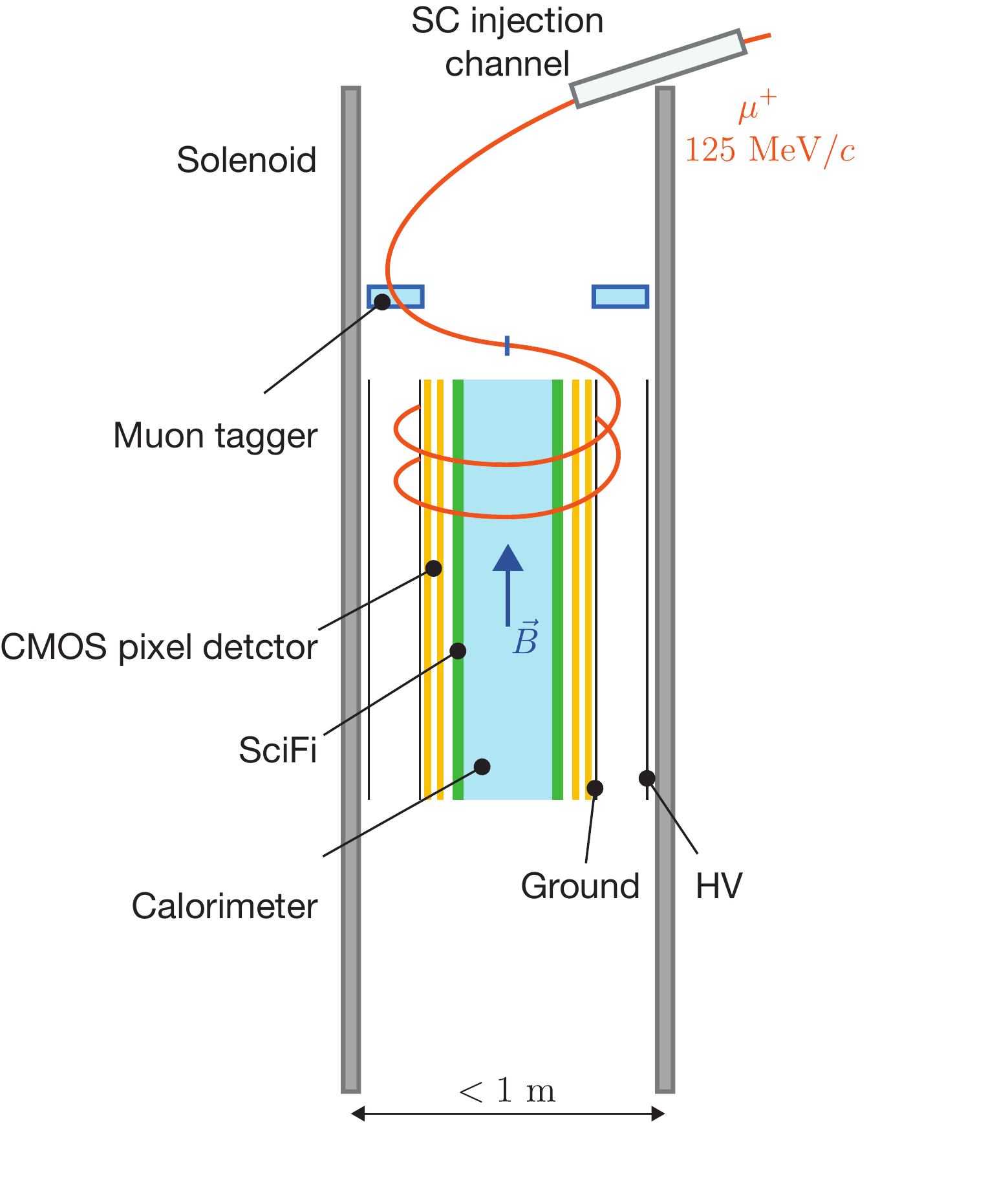}
\caption{Left: Lateral muEDM, featuring a horizontal beam injection into a ring-shaped compact storage ring. Right: Helix muEDM, featuring a vertical beam injection into a compact solenoid.}
\label{fig:muEDM}
\end{figure}

Two concepts have been studied for a dedicated muon EDM search using the frozen-spin technique~\cite{Adelmann:2021udj}, namely a \emph{Lateral} muEDM and a \emph{Helix} muEDM. The Lateral muEDM approach has been promoted by Adelmann \textit{et al.}~\cite{Adelmann:2010zz}: a muon beam is injected laterally into a ring-shaped compact storage ring, as depicted in Fig.~\ref{fig:muEDM} (left). One of the downsides of this configuration is the deterioration of the beam quality due to the multiple Coulomb scattering on the electrodes of the radial electric field. This results in a lower storage efficiency. Another challenge is to develop a fast trigger, of the order of ns, to initiate the beam injection and the storage process. The Helix muEDM approach, on the other hand, is promoted and inspired by the J-PARC muon $g-2$/EDM experiment, where beam injection under a vertical angle is employed~\cite{Iinuma:2016zfu}. The clear advantage is that the muons do not have to pass several times through electrodes, as in the Lateral muEDM approach. Moreover, the requirement of triggering the injection process relaxes to about 50\,ns. A sketch of the apparatus for the Helix muEDM is depicted in Fig.~\ref{fig:muEDM} (right). By using a $125\,\mathrm{MeV}/c$ muon beam with 95\% polarization, a radial electric field of $E_{r}=2$\,MV/m, and a magnetic field of 3\,T, the projected sensitivity is $6 \times 10^{-23}\,e\cdot$cm after one year of data taking, assuming $N \sim 10^{12}$.

\section{Current status of the activities at PSI}

Two test-beam periods have been accomplished at PSI; in 2019, the main purpose was to characterize the phase space of the potential muon beamlines; in 2020, the main motivation was to study the effect of multiple Coulomb scattering of positrons at low momentum. For the measurement at $\pi$E1, the rate is up to $4.4 \times 10^6$ $\mu^{+}$/s, for 1.6\,mA proton current, for muons with $p=28\,\mathrm{MeV}/c$, and emittances $\epsilon_{x}=197.7$\,mm~mrad and $\epsilon_{y}=171.3$\,mm~mrad. The $\mu$E1 beamline, on the other hand, has a rate of up to $8 \times 10^7$ $\mu^{+}$/s, for 1.6\,mA proton current, for muons of $p=125\,\mathrm{MeV}/c$, and emittances  $\epsilon_{x}=945$\,mm~mrad and $\epsilon_{y}=716$\,mm~mrad. The measured polarization is $\sim$95\%. The $\pi$E1 beamline is a candidate for a precursor experiment at low magnetic field (0.7\,T), while the $\mu$E1 beamline is the candidate for the final data-taking using a 3-T magnetic field. 

\begin{figure}[htbp]
\centering
\includegraphics[width=0.35\textwidth]{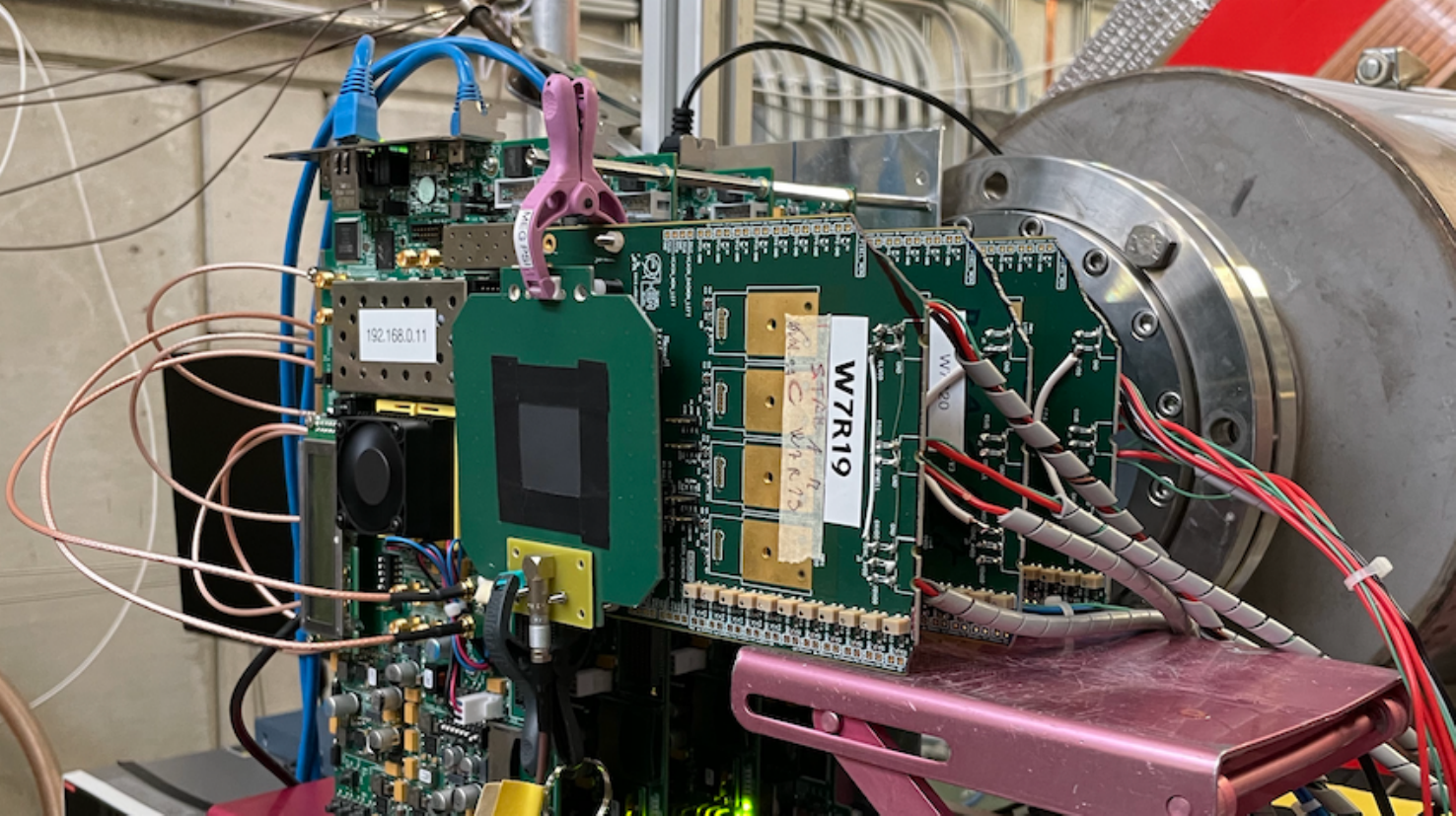}
\includegraphics[width=0.35\textwidth]{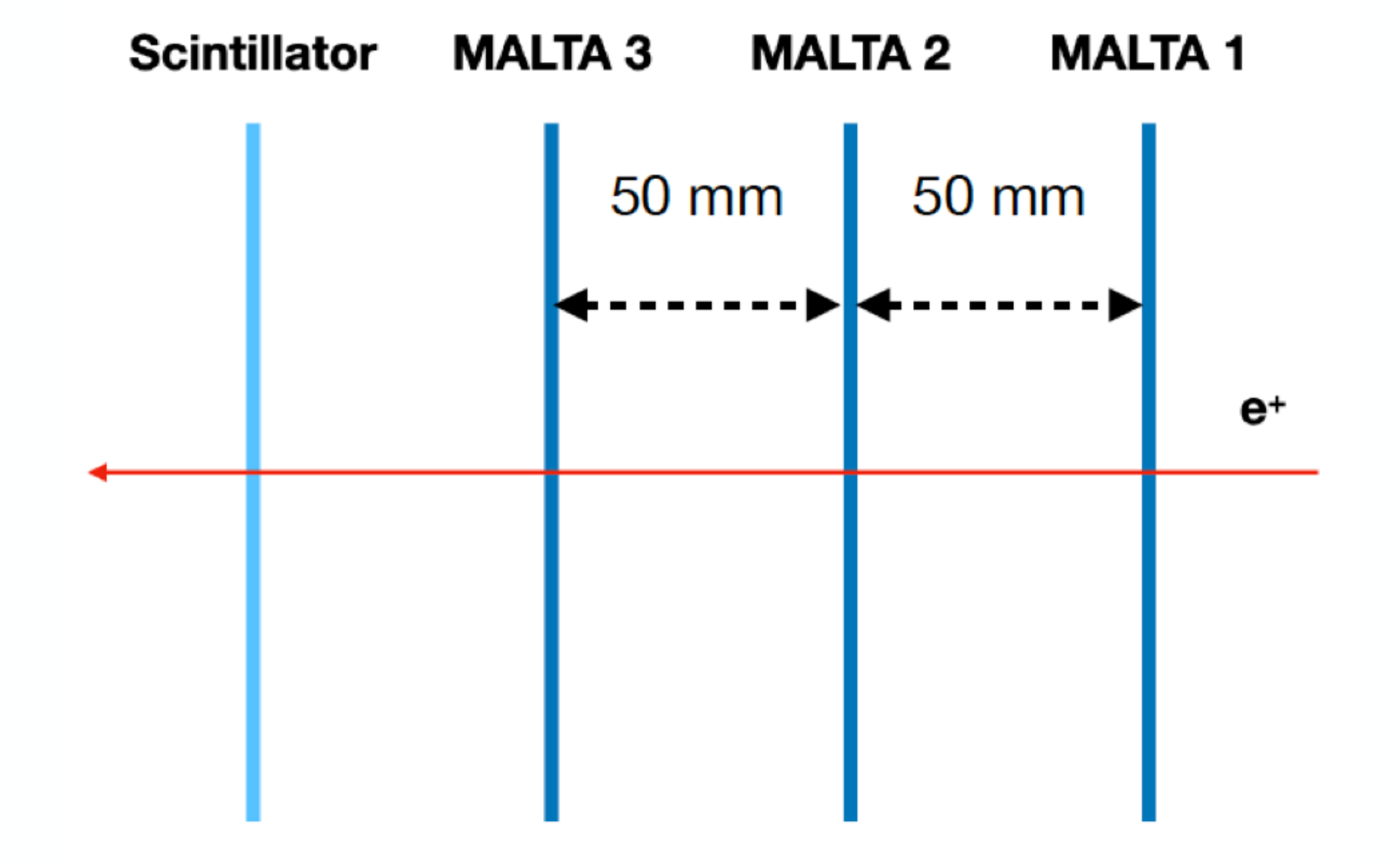}
\caption{Picture (left) and sketch (right) of the test beam apparatus installed at PSI in 2020.}
\label{fig:malta}
\end{figure}

The multiple Coulomb scattering study was performed using the 3-plane MALTA telescope~\cite{Cardella:2019ksc}. Each plane has a matrix of $512 \times 512$\,pixels, and each pixel has an active area of $36.4 \times 36.4$\,$\mu$m$^{2}$. The sensor is made of 300\,$\mu$m-thick silicon. The momentum range of positron studied was 20 - 85\,MeV/$c$. Two configurations were studied: $i)$ MALTA as active target; $ii)$ MALTA + Kapton/Mylar. The obtained RMS value of the multiple Coulomb scattering angle per layer agrees well with the Highland formula~\cite{Highland:1975pq}. 

\section{Conclusions and outlook}

The search for the muon EDM will provide an additional handle to BSM physics in the muon sector. The frozen-spin technique can provide a more sensitive search compared to the approach utilizing the muon $g-2$ storage ring. A sensitivity of $6\times10^{-23}~e\cdot$cm can be reached using existing beamlines at PSI. R\&D of key components of the experimental apparatus is currently underway for a full experimental proposal to be submitted to PSI CHRISP research committee soon.

\section*{Acknowledgments}

We would like to thank H. Pernegger and C. Solans Sanchez for their support and for providing the MALTA sensors for multiple scattering measurement. We would also like to thank M. Hoferichter for a very useful discussion regarding the SM prediction of the muon EDM. This work is supported by the National Natural Science Foundation of China under Grant No. 12075151 and 12050410233 and the ETH Research Grant ETH-48 18-1.

\end{document}